\documentclass{article}
\usepackage{amsfonts}
\usepackage{amssymb}
\usepackage{amsmath}

\newtheorem{theorem}{Theorem}

\newtheorem{proposition}[theorem]{Proposition}

\input{tcilatex}
\begin{document}

\begin{center}
{\Large Poisson Structures for Aristotelian Model of}

\bigskip

{\Large Three Body Motion}

\bigskip

\bigskip E. Abado\u{g}lu and H. G\"{u}mral \ 

Department of Mathematics, Yeditepe University,

34755 Ata\c{s}ehir, \.{I}stanbul, Turkey

eabadoglu@yeditepe.edu.tr,\ \ hgumral@yeditepe.edu.tr
\end{center}

\bigskip

\bigskip

\textbf{Abstract: }We present explicitly Poisson structures, for both
time-dependent and time-independent Hamiltonians, of a dynamical system with
three degrees of freedom introduced and studied by Calogero \textit{et al }%
[2005]\textit{.} For the time-independent case, new constant of motion
includes all parameters of the system. This extends the result of Calogero 
\textit{et al }[2009] for semi-symmetrical motion. We also discuss the case
of three bodies two of which are not interacting with each other but are
coupled with the interaction of third one.

\bigskip

\section{Introduction}

In references \cite{cal05},\cite{cal08} Calogero \textit{et al. }suggested
and studied the system of differential equations

\begin{eqnarray}
\dot{x} &=&i\omega x+\frac{c}{x-y}+\frac{b}{x-z}  \notag \\
\dot{y} &=&i\omega y+\frac{a}{y-z}+\frac{c}{y-x}  \label{c1} \\
\dot{z} &=&i\omega z+\frac{b}{z-x}+\frac{a}{z-y}  \notag
\end{eqnarray}
as a model for motions of three bodies where forces determine velocities
(hence the name Aristotelian), rather than accelerations . They were
considered to be a prototype for a large class of models exhibiting
transitions from simple to complicated motions that can be explained as
travel on Riemann surfaces (see also extensive list of references in \cite%
{cal05} and the fortcoming article \cite{cal11}). In this work, following
references \cite{hg99} and \cite{hasan}, we shall present Poisson structures
of an equivalent system of differential equations obtained after elimination
of the linear one-body forces in Eq.(\ref{c1}).

Differentiation of Eq.(\ref{c1}) implies that they may be interpreted as
Newtonian equations for an inverse-cubic force field. Historically, cubic
forces were first considered, in the Newtonian context, by Jacobi \cite%
{jac66}. Poincar\'{e}, in his search of infinitely many periodic solutions
of three body problem, replaced the inverse-square force with an
inverse-cubic one \cite{poi96}. The relation of inverse-cubic forces with
integrable three body problem of Newtonian type with inverse-square
potential were explained in an appendix of reference \cite{cal08}. See also 
\cite{cal69}-\cite{cal08b} for other treatments of inverse-cubic forces.

The system in Eq.(\ref{c1}) was communicated to us by Yavuz Nutku who also
presented its time-dependent conserved Hamiltonian functions \cite{yn00}
(c.f. Eqs.(\ref{h1})-(\ref{hh2})). A transformation of variables (c.f. Eq.(%
\ref{tr})), first appeared in reference \cite{cal05}, takes one of the
Hamiltonian functions into time-independent form but the second one still
remains time-dependent. Using techniques of reference \cite{hg99}, we shall
present, in section (3), a formal Hamiltonian structure with this
time-dependent Hamiltonian function.

We shall indicate, in section (4), a potential function by which the
dynamical equations can be cast into a gradient flow. A linear change of
variables that includes time-independent Hamiltonian function as one of the
new coordinates reduces the number of equations into two which are still in
gradient form. We shall then be able to integrate for the second
time-independent conserved Hamiltonian. For the full-symmetrical and
semi-symmetrical cases, characterized by $a=b=c$ and $a=b\neq c$,
respectively, the bi-Hamiltonian form of equations for Aristotelian model of
three body motion will be presented in section (5). It becomes necessary to
give a separate treatment of a subcase of the semi-symmetrical motion
characterized by $a=b\neq 0$ and $c=0$. This correponds to a model of three
body motion in which two of the bodies do not interact with each other, yet
their motions are coupled due to their interaction with the third body.

In section (6), we shall construct only the second time-independent
Hamiltonian function containing all coupling constant, which is enough to
cast the system into bi-Hamiltonian form in three-dimensions. We shall first
complete the discussion for non-interacting two body case by considering $%
a\neq b\neq 0$ and $c=0$. We shall then present the general form of
Hamiltonian function for generic values of coupling constants. This
generalizes a result of reference \cite{cal08} where a time-independent
conserved function was found by a different analysis for the
semi-symmetrical case. Present results are contributions with a geometric
taste to the ongoing investigations \cite{cal11} of the authors of
references \cite{cal05} and \cite{cal08} where they have been presenting
full analysis of the dynamics such as methods of integrating Eq.(\ref{c1}),
analytic structure of solutions, as well as the geometry of the Riemann
surfaces involved.

\section{Aristotelian Model of 3-Body Motion \label{nutku}}

In \cite{cal05}, the motion of three point particles, for which the
velocities, rather than accelarations, are determined by forces, was
modelled by the equations 
\begin{equation}
\dot{z}_{n}=i\omega z_{n}+\frac{g_{n-2}}{z_{n}-z_{n-1}}+\frac{g_{n-1}}{%
z_{n}-z_{n-2}}  \label{zeq}
\end{equation}%
where $g_{n}$ are coupling constants, $z_{n}$ are (complex) coordinates of
particle positions, $n=1,2,3$, $\omega >0$, and $i=\sqrt{-1}$. Overdots
denote derivatives with respect to (real) time variable $t$ and Eq.(\ref{zeq}%
) was referred to as physical model. It was remarked that this systems
admits canonical Hamiltonian formalism with the Hamiltonian function

\begin{equation}
H(z,p)=\Sigma _{n=1}^{3}[-i\omega z_{n}p_{n}+g_{n}\frac{p_{n+1}-p_{n+2}}{%
z_{n+1}-z_{n+2}}]
\end{equation}%
where the momenta $p_{n}$ look like Lagrange multipliers.

We shall consider the system of equations (\ref{zeq}), and not its extension
by the addition of momenta, with $z_{1}=x,\;z_{2}=y,\;z_{3}=z$ and $%
g_{1}=a,\,g_{2}=b,\,g_{3}=c$ resulting in Eq.(\ref{c1}). Nutku \cite{yn00}
indicated two time-dependent conserved Hamiltonian functions 
\begin{eqnarray}
H^{(1)}(x,y,z) &=&e^{-i\omega t}(x+y+z)  \label{h1} \\
H^{(2)}(x,y,z) &=&\frac{1}{4}e^{-4i\omega t}(x^{2}+y^{2}+z^{2})-(a+b+c)t
\label{h2}
\end{eqnarray}%
for Eq.(\ref{c1}). The choice of variables, first appeared in \cite{cal05}, 
\begin{equation}
(u,v,w,\tau )=e^{-i\omega t}(x,y,z,-\frac{e^{-i\omega t}}{2i\omega })
\label{tr}
\end{equation}%
factors out the linear forces and transforms the system in Eq.(\ref{c1})
into the one with two-body interactions only 
\begin{eqnarray}
\acute{u} &=&\frac{c}{u-v}+\frac{b}{u-w}  \notag \\
\acute{v} &=&\frac{a}{v-w}+\frac{c}{v-u}  \label{c2} \\
\acute{w} &=&\frac{b}{w-u}+\frac{a}{w-v}.  \notag
\end{eqnarray}%
Here, prime denotes derivative with respect to complexified time $\tau $
and, for this reason Eq.(\ref{c2}) was referred to as auxiliary model \cite%
{cal05}. This transformation makes the first Hamiltonian function $H^{(1)}$
independent of time, but nevertheless $H^{(2)}$ is still time-dependent%
\begin{eqnarray}
H^{(1)}(u,v,w) &=&u+v+w  \label{hh1} \\
H^{(2)}(u,v,w) &=&u^{2}+v^{2}+w^{2}-2(a+b+c)\tau \text{.}  \label{hh2}
\end{eqnarray}%
These relations were used in \cite{cal08} to obtain the general solution of
the system (\ref{c2}). Level sets of $H^{(1)}$ are planes and of $H^{(2)}$
are spheres expanding with velocity $2(a+b+c)/\sqrt{2(a+b+c)\tau +H_{0}^{(2)}%
}$ where $H_{0}^{(2)}$ is the initial radius of the sphere. $H^{(1)}$
represents the coordinates of center of mass of three bodies. We note that%
\begin{equation*}
H^{(3)}(u,v,w)=uv+uw+vw+(a+b+c)\tau
\end{equation*}%
is also a time-dependent conserved function for Eq.(\ref{c2}), but it is not
independent from $H^{(1)}$ and $H^{(2)}$ 
\begin{equation*}
H^{(3)}=\frac{1}{2}((H^{(1)})^{2}-H^{(2)}).
\end{equation*}%
Moreover, there is another relation between the time-dependent Hamiltonians%
\begin{equation*}
(u-v)^{2}+(v-w)^{2}+(w-u)^{2}=2H^{(2)}-2H^{(3)}+4(a+b+c)\tau
\end{equation*}%
which can be used to eliminate the time variable $\tau $. This seemed to be
useful in obtaining second time-independent Hamiltonian function in somewhat
different analysis of reference \cite{cal08}.

Apart from the semi-symmetrical and the full-symmetrical (or integrable)
configurations, the two-body case was obtained by the restriction $a=b=0$
and $c\neq 0$ \cite{cal05} which is not much interesting in the present
framework. However, in presenting bi-Hamiltonian structures, we shall treat
separately a particular case of semi-symmetrical motion in which the
interaction between two of the bodies is neglected. Note also that, by a
rescaling of time $\tau $ one of the coupling constants can be removed from
the right hand side of equations (\ref{c2}). In this work, we shall consider
the variables $\mathbf{u}=(u,v,w)$ in a real domain. In the sequel $U=%
\mathbf{U}\cdot \nabla =\mathbf{\acute{u}}\cdot \nabla $ will denote the
vector field associated with Eq.(\ref{c2}).

\section{Time-dependent Poisson structures}

We first present the Hamiltonian structure of equations (\ref{c2}) with the
time-dependent Hamiltonian function $H^{(2)}$. This can be achieved by
considering Poisson structures in space-time variables $(\tau ,\mathbf{u})$.
In other words, we add the equation $\acute{\tau}=1$ to the system in Eq.(%
\ref{c2}) (if necessary, after performing a linear change of parametrization
by time). The flow of suspended system in four dimensions is then generated
by the vector field $\partial _{t}+U$ and our task is to find a Hamiltonian
representation of this. We shall use contravariant description of
Hamiltonian formalism by Poisson structures.

A Poisson structure on a manifold $N$ is defined by a skew symmetric
contravariant bilinear form subjected to the Jacobi identity expressed as
the vanishing of Schouten bracket of Poisson tensor with itself \cite{lich77}%
-\cite{mr94}. Following \cite{hg99}, for a Hamiltonian formalism on a
time-extended space $N=I\times M$, $I\subset 
\mathbb{R}
$ or $%
\mathbb{C}
$ and $M\subset 
\mathbb{R}
^{3}$ for the present context, we take the bi-vector field \ 
\begin{equation}
\Lambda (\tau ,\mathbf{u})=V(\tau ,\mathbf{u})\wedge \partial _{\tau }+\Pi
(\tau ,\mathbf{u})  \label{extp}
\end{equation}%
where $V$ and $\Pi $ are time-dependent vector and bi-vector fields on $M$,
respectively. The Jacobi identity for $\Lambda $ can be computed in a
coordinate independent way using identities of Schouten algebra of
multi-vectors.

\begin{proposition}
$\Lambda $ is a Poisson bi-vector field on $I\times M$ if and only if%
\begin{equation}
\lbrack \Pi ,\Pi ]=2V\wedge \frac{\partial \Pi }{\partial \tau }\text{, \ \
\ \ \ \ }[\Pi ,V]=V\wedge \frac{\partial V}{\partial \tau }.  \label{jacobi}
\end{equation}
\end{proposition}

The Hamiltonian form of the suspended vector field $\partial _{t}+U$ on $%
I\times M$ will then be 
\begin{equation}
\partial _{t}+U=\Lambda (dH)=V(H)\partial _{t}+\Pi (dH)-H_{,t}V  \label{eqm}
\end{equation}%
where $H$ is a time-dependent conserved function of $U$. Due to
non-linearity of Jacobi identity, finding bi-vector $\Lambda $ from Eqs.(\ref%
{jacobi}) and (\ref{eqm}) for given Hamiltonian function $H$, is a difficult
task. The next result, which is the linearization of Jacobi identity by
Hamiltonian vector fields, may help.

\begin{proposition}
For the Hamiltonian vector field in Eq.(\ref{eqm}), the Poisson bi-vector (%
\ref{extp}) satisfies the infinitesimal invariance conditions%
\begin{equation}
\frac{\partial V}{\partial \tau }+[U,V]=0,\text{ \ \ \ \ \ \ }\frac{\partial
\Pi }{\partial \tau }+[U,\Pi ]=V\wedge \frac{\partial U}{\partial \tau }
\label{eje}
\end{equation}%
which are equivalent to the Jacobi identity in Eq.(\ref{jacobi}). Moreover, $%
V$ is an infinitesimal automorphism of $\Lambda $.
\end{proposition}

This follows from the Jacobi identity (of Schouten algebra) for
multi-vectors $(\Lambda ,\Lambda ,h)$. Conversely, one obtains Eq.(\ref%
{jacobi}) by inserting Hamilton's equations (\ref{eqm}) into Eq.(\ref{eje}).
The last conclusion follows from second of Eq.(\ref{jacobi}).

Thus, the construction of Hamiltonian structure of a system admitting
time-dependent Hamiltonian function amounts to solving the linear system
consisting of Eqs.(\ref{eqm}), (\ref{eje}) and the conservation law for $H$.
As it can be inferred from Eq.(\ref{eqm}) the function $H$ is, in addition,
coupled to the vector field $V$ by the condition $V(H)=1$.

The construction of Hamiltonian structure for the Aristotelian model of
dynamical equations relies on the observation that the first of Eq.(\ref{eje}%
) gives a characterization of the vector field $V$ as a time-dependent
infinitesimal symmetry of $U$. Denoting%
\begin{equation*}
E=u\partial _{u}+v\partial _{v}+w\partial _{w}
\end{equation*}%
the Euler vector field and, noting that $U$ is homogeneous of degree $-1$,
it is easy to see that the vector field 
\begin{equation*}
E-2\tau U=(u-\frac{2\tau c}{u-v}-\frac{2\tau b}{u-w})\partial _{u}+(v-\frac{%
2\tau a}{v-w}-\frac{2\tau c}{v-u})\partial _{v}+(w-\frac{2\tau b}{w-u}-\frac{%
2\tau a}{w-v})\partial _{w}
\end{equation*}%
is a time-dependent infinitesimal symmetry of $U$, that is, 
\begin{equation*}
\lbrack \partial _{\tau }+U,E-2\tau U]=0.
\end{equation*}%
This condition is essential to show that the bi-vector field 
\begin{equation}
\Lambda =(E-2\tau U)\wedge \partial _{\tau }+E\wedge U  \label{poi}
\end{equation}%
is Poisson $[\Lambda ,\Lambda ]=0$. We can then cast the system (\ref{c2})
with time-dependent conserved function in Eq.(\ref{hh2}) into an autonomous
Hamiltonian system in four-dimensions.

\begin{proposition}
The vector field $(1,\mathbf{U})$ is Hamiltonian for the Poisson bi-vector
in Eq.(\ref{poi}) and with the Hamiltonian function defined by the
time-dependent function in Eq.(\ref{hh2}) 
\begin{equation*}
\partial _{\tau }+U=\Lambda (dH)\text{, \ \ \ \ \ }H=\frac{1}{2}\ln H^{(2)}.
\end{equation*}
\end{proposition}

This rather formal Hamiltonian structure may be useful in investigation of
geometric structure of solution space as well as symmetries and invariants
of the flow thereon \cite{hg99}. For example, it follows that the
time-dependent infinitesimal symmetry $E-2\tau U$ of $U$ is a Hamiltonian
vector field for $\Lambda $ with the Hamiltonian function $\tau $. In other
words, the solution space of Eq.(\ref{c2}) may be realized as level sets of
Hamiltonian function of an infinitesimal symmetry of motion it describes.
The conserved function $H^{(1)}$ gives the Hamiltonian vector field $%
H^{(1)}(\partial _{\tau }+U)$ whereas $H^{(3)}$ results in $-2H^{(3)}U$.

\section{Potential Function}

The force field of Aristotelian model can be derivable from a logarithmic
potential field.

\begin{proposition}
The dynamical vector field $U$ in Eq.(\ref{c2}) generating the Aristotelian
motion of three bodies is a gradient field $\mathbf{U=}\nabla F$, with the
potential function 
\begin{equation*}
F=\ln (v-w)^{a}(u-w)^{b}(u-v)^{c}.
\end{equation*}
\end{proposition}

Note that the potential function $F$ is a solution of the partial
differential equation%
\begin{equation*}
\partial _{u}F+\partial _{v}F+\partial _{w}F=0
\end{equation*}%
which is another manifestation of the conservation law for $H^{(1)}$.

We shall consider a linear transformation of Cartesian coordinates $(u,v,w)$
by which the conserved Hamiltonian function $H^{(1)}$ is eliminated and the
potential function $F$ becomes a function of two variables. The reduced
system in two variables, which is still a gradient flow, admits a
non-canonical symplectic structure similar to the one for the Lotka-Volterra
system in \cite{nutku}.

\begin{proposition}
In the orthonormal coordinates%
\begin{equation}
\zeta =\frac{1}{\sqrt{3}}(u+v+w)\text{, \ }\eta =\frac{1}{\sqrt{2}}(u-v)%
\text{, \ \ }\xi =\frac{1}{\sqrt{6}}(u+v-2w)  \label{newcoor}
\end{equation}%
the potential function $F$ becomes a function of $(\eta ,\xi )$ only 
\begin{equation}
F(\eta ,\xi )=\ln \eta ^{c}(\sqrt{3}\xi +\eta )^{b}(\sqrt{3}\xi -\eta )^{a}%
\text{.}  \label{pot}
\end{equation}%
\ 
\end{proposition}

In the new coordinates the right handed orthonormal basis vectors are%
\begin{equation*}
\mathbf{e}_{1}=\nabla \zeta \text{, \ \ }\mathbf{e}_{2}=\nabla \eta \text{,
\ \ }\mathbf{e}_{3}=\nabla \xi .
\end{equation*}%
From the inversion of Eq.(\ref{newcoor}) we observe that the combinations%
\begin{equation*}
u-v=\sqrt{2}\eta ,\ \ \ v-w=\frac{1}{\sqrt{2}}(\sqrt{3}\xi -\eta ),\ \ \
w-u=-\frac{1}{\sqrt{2}}(\sqrt{3}\xi +\eta )
\end{equation*}%
are independent of $\zeta =H^{(1)}/\sqrt{3}$.

\begin{proposition}
On level sets of Hamiltonian function $H^{(1)}$ with coordinates $(\eta ,\xi
)$, the dynamical system in Eq.(\ref{c2}) becomes%
\begin{equation}
\text{\ }\acute{\eta}=\frac{c}{\eta }+\frac{b}{\sqrt{3}\xi +\eta }-\frac{a}{%
\sqrt{3}\xi -\eta }\text{, \ \ \ }\acute{\xi}=\sqrt{3}(\frac{b}{\sqrt{3}\xi
+\eta }+\frac{a}{\sqrt{3}\xi -\eta })  \label{c3}
\end{equation}%
which is also a gradient system with the potential in Eq.(\ref{pot}).
\end{proposition}

These are the reduced equations of motion after elimination of the motion of
center of mass. General theory of reduction implies that the reduced space
must be symplectic \cite{mr94}. In fact, as any orientable two dimensional
space is symplectic, we can take an area form $\phi (\eta ,\xi )d\xi \wedge
d\eta $ as the symplectic two-form for Eq.(\ref{c3}). In the context of
local structure of Poisson manifolds, as described in reference \cite%
{weinstein}, this reduction gives the symplectic foliation of the three
dimensional space of variables $(u,v,w)$. Liouville theorem for the adapted
symplectic form and Eq.(\ref{c3}) implies the invariance of the volume
elements in the sense that the volume density $\phi (\eta ,\xi )$ satisfies 
\begin{equation*}
\acute{\eta}\frac{\partial \phi }{\partial \eta }+\acute{\xi}\frac{\partial
\phi }{\partial \xi }+\phi \nabla ^{2}F(\eta ,\xi )=0
\end{equation*}%
where $\nabla $ denotes $(\partial _{\eta },\partial _{\xi })$. The
characteristics of this equation is the second time-independent conserved
Hamiltonian we are looking for. It is also the Hamiltonian function for Eq.(%
\ref{c3}) with respect to the symplectic structure introduced above.

\section{Bi-Hamiltonian structures in $%
\mathbb{R}
^{3}$}

The construction of bi-Hamiltonian structure in three dimensions requires
two (time-independent) Hamiltonian functions $H_{1}$, $H_{2}$ and a
conformal factor $\phi $ (this will be seen to be the same function
involving symplectic structure mentioned above) such that Hamilton's
equations take the form%
\begin{equation*}
\mathbf{\acute{u}}=\phi \nabla H_{1}\times \nabla H_{2}.
\end{equation*}%
The Poisson tensors can be identified with vectors $\phi \nabla H_{1}$ and $%
-\phi \nabla H_{2}$ using the isomorhism between three vectors in $%
\mathbb{R}
^{3}$ and $3\times 3$ skew-symmetric matrices. The corresponding Hamiltonian
functions are $H_{2}$ and $H_{1}$, respectively. Poisson tensors constructed
this way can always be made into a compatible pair. \textit{A priori}
unspecified function $\phi $ is related to an invariance property of the
Jacobi identity in three dimensions. Namely, any multiple of a Poisson
tensor with an arbitrary function is also a Poisson tensor. Thus, to cast a
system into bi-Hamiltonian form, we need to find fundamental conserved
quantities and then determine the multiplicative function $\phi $. See
references \cite{hasan},\cite{nambu}-\cite{aygur} for more details and
various explicit examples.

Since we have the time-independent conserved Hamiltonian $H^{(1)}$ given by
Eq.(\ref{hh1}), our first task is to find a second one. We want to search
for the second time-independent conserved quantity for the system in Eq.(\ref%
{c2}) or, equivalently, in Eq.(\ref{c3}) using the reduced gradient flow of
the latter. Note that, if there exist a time-independent conserved
Hamiltonian function $H(\eta ,\xi )$, then we can write%
\begin{equation*}
\nabla F=\phi \nabla \zeta \times \nabla H=\mathbf{e}_{1}\times \phi \nabla H
\end{equation*}%
for some function $\phi $. It then follows that the non-zero part of Eq.(\ref%
{c3}) admit (noncanonical) symplectic formulation in the $(\eta ,\xi )$%
-variables%
\begin{equation}
\partial _{\eta }F=-\phi \partial _{\xi }H\text{, \ \ \ \ }\partial _{\xi
}F=\phi \partial _{\eta }H  \label{symp}
\end{equation}%
with the symplectic two-form $\phi (\eta ,\xi )d\xi \wedge d\eta $ which is
always closed and non-degenerate in two dimensions. Thus, the linear change
of coordinates in Eq.(\ref{newcoor}) enables us to realize the local
structure of Poisson manifold that we are going to construct. More
precisely, the symplectic foliation of the space of variables $(\zeta ,\eta
,\xi )$ consists of coordinate planes $\zeta =$constant, or equivalently,
the level sets of the conserved Hamiltonian function $H^{(1)}$.

The Hamiltonian function $H(\eta ,\xi )$ will be a function of
characteristic curves defined by the Hamiltonian system in Eq.(\ref{symp})
and it may depend arbitrarily on the variable $\zeta $. To find the
characteristic curves of Eq.(\ref{symp}), or what we shall call the
fundamental conserved quantity, we eliminate the time derivatives in Eq.(\ref%
{c3}) and obtain the ordinary differential equation%
\begin{equation}
\frac{d\xi }{d\eta }=-\frac{\sqrt{3}(a-b)\eta ^{2}+3(a+b)\eta \xi }{%
(a+b+c)\eta ^{2}+\sqrt{3}(a-b)\eta \xi -3c\xi ^{2}}  \label{de}
\end{equation}%
which is homogeneous of degree two. So, for $\theta =\xi /\eta ,$ its
solution defines the characteristics 
\begin{equation}
\text{constant}=\ln \eta +\int \frac{(a+b+c)+\sqrt{3}(a-b)\theta -3c\theta
^{2}}{\sqrt{3}(a-b)+(4a+4b+c)\theta +\sqrt{3}(a-b)\theta ^{2}-3c\theta ^{3}}%
d\theta   \label{integ}
\end{equation}%
or, equivalently, the fundamental conserved quantity for the flow. That
means, the time-independent Hamiltonian function we are seeking is a
function of $\zeta $ and the function defined by Eq.(\ref{integ}). Starting
from the simplest, we shall first present explicitly bi-Hamiltonian
structures of symmetrical cases and then proceed, in the next section, to
analyse the generic solution for the fundamental conserved quantity in Eq.(%
\ref{integ}). We shall see that, in semi-symmetrical motion, the subcase $c=0
$ requires a separate treatment and we shall do this in the last subsection.

\subsection{Full-symmetrical case}

We set $a=b=c$ for which Eq.(\ref{de}) becomes independent of coupling
constants%
\begin{equation}
\frac{d\xi }{d\eta }=-\frac{2\eta \xi }{\eta ^{2}-\xi ^{2}}
\end{equation}%
and the solution gives%
\begin{equation}
h_{f}(\xi ,\eta )=\xi ^{3}-3\xi \eta ^{2}  \label{fullham}
\end{equation}%
which can easily be verified to be conserved under the flow of Eq.(\ref{c3}%
). Simple manipulations show that%
\begin{equation*}
\phi _{f}(\xi ,\eta )=1/\sqrt{3}\eta (3\xi ^{2}-\eta ^{2})
\end{equation*}%
is the conformal factor entering the definition of symplectic structure. As
a function of variables $(u,v,w)$ in Eq.(\ref{c2}), we find

\begin{proposition}
For the Aristotelian model with equal coupling constants, the dynamical
equations%
\begin{eqnarray}
\acute{u} &=&\frac{1}{u-v}+\frac{1}{u-w}  \notag \\
\acute{v} &=&\frac{1}{v-w}+\frac{1}{v-u}  \label{full} \\
\acute{w} &=&\frac{1}{w-u}+\frac{1}{w-v}  \notag
\end{eqnarray}%
admit bi-Hamiltonian structure 
\begin{equation*}
\mathbf{\acute{u}}=P_{f1}(\mathbf{u})\nabla H_{f}(\mathbf{u})=P_{f2}(\mathbf{%
u})\nabla H^{(1)}(\mathbf{u})
\end{equation*}%
with the following pairs of Poisson matrices and Hamiltonian functions%
\begin{eqnarray*}
P_{f1}(\mathbf{u}) &=&\frac{-1}{\sqrt{6}(u-v)(v-w)(w-u)}\left( 
\begin{array}{ccc}
0 & 1 & -1 \\ 
-1 & 0 & 1 \\ 
1 & -1 & 0%
\end{array}%
\right) \text{, \ \ \ \ \ \ } \\
H_{f}(\mathbf{u}) &=&\frac{1}{6\sqrt{6}}(u+v-2w)[(u+v-2w)^{2}-9(u-v)^{2}], \\
P_{f2}(\mathbf{u}) &=&\text{\ }\frac{1}{6}[\frac{(u-v)}{(v-w)(w-u)}+\frac{2}{%
u-v}]\left( 
\begin{array}{ccc}
0 & -2 & -1 \\ 
2 & 0 & 1 \\ 
1 & -1 & 0%
\end{array}%
\right)  \\
&&+\frac{1}{2}[\frac{1}{v-w}-\frac{1}{w-u}]\left( 
\begin{array}{ccc}
0 & 0 & 1 \\ 
0 & 0 & 1 \\ 
-1 & -1 & 0%
\end{array}%
\right) , \\
H^{(1)}(\mathbf{u}) &=&u+v+w.
\end{eqnarray*}
\end{proposition}

The conformal factor for these Poisson structures is the function $\phi _{f}(%
\mathbf{u})$ given as the coefficient of the constant matrix of $P_{f1}(%
\mathbf{u})$. This function constitutes an invariant volume density for the
solution space of variables $\mathbf{u}$ in the sense that the three-form $%
\phi _{f}(\mathbf{u})du\wedge dv\wedge dw$ has vanishing Lie derivative with
respect to the vector field defined by the right hand side of Eq.(\ref{full}%
).

\subsection{Semi-symmetrical case}

Setting $a=b\neq c$, we have 
\begin{equation}
\frac{d\xi }{d\eta }=-\frac{6a\eta \xi }{(2a+c)\eta ^{2}-3c\xi ^{2}}
\end{equation}%
and its integration gives%
\begin{equation}
\eta (\frac{\xi }{\eta })^{\mu }[(\frac{\xi }{\eta })^{2}-\frac{1}{4\mu -1}%
]^{(1-\mu )/2}=\text{constant}  \label{char}
\end{equation}%
where, following reference \cite{cal08}, we introduce the constant $\mu
=(2a+c)/(8a+c)$. Various powers of the function in Eq.(\ref{char}) can be
adapted as the fundamental conserved quantity. In any case, we have to
exclude the values $\mu =1/4$ (or $c=0$) and $\mu =1$ (or $a=0$) in the
following discussions. For $a=0$, the evolution in the variable $w$
disappears (c.f. Eq.(\ref{ss})) and we are left with a two dimensional
system. For $c=0$, we have the special case of three body motion in which
two of the bodies do not interact with each other. This necessarily requires
separate treatment which we will take up in the next subsection. We refer to
extensive discussion in reference \cite{cal08} where it was shown that
values of the constant $\mu $, in particular, the real rational values, play
important role in determining dynamical evolution of the model.

For the purposes of having a generalization of the full-symmetrical case and
the same invariant volume density, we choose the fundamental conserved
quantity for semi-symmetrical case to be%
\begin{equation*}
h_{s}(\xi ,\eta )=\xi ^{2\mu /(1-\mu )}[\xi ^{2}-\frac{1}{4\mu -1}\eta ^{2}]%
\text{, \ \ \ \ \ }\mu \neq 1,1/4
\end{equation*}%
so that, when $a=c$, or equivalently, $\mu =1/3$ we obtain the function $%
h_{f}(\xi ,\eta )$ in Eq.(\ref{fullham}).

\begin{proposition}
For the Aristotelian model with two coupling constants $a\neq 0$ and $c\neq 0
$, the dynamical equations%
\begin{eqnarray}
\acute{u} &=&\frac{c}{u-v}+\frac{a}{u-w}  \notag \\
\acute{v} &=&\frac{a}{v-w}+\frac{c}{v-u}  \label{ss} \\
\acute{w} &=&\frac{a}{w-u}+\frac{a}{w-v}  \notag
\end{eqnarray}%
admit Hamiltonian structure with the Poisson tensor 
\begin{equation*}
P_{s1}(\mathbf{u})=-\frac{3c}{2}\frac{(u+v-2w)^{(1-3\mu )/(1-\mu )}}{%
(u-v)(v-w)(w-u)}\left( 
\begin{array}{ccc}
0 & 1 & -1 \\ 
-1 & 0 & 1 \\ 
1 & -1 & 0%
\end{array}%
\right) \text{, \ \ \ }
\end{equation*}%
and with the Hamiltonian function%
\begin{equation*}
H_{s}(\mathbf{u})=(u+v-2w)^{2\mu /(1-\mu )}[(u+v-2w)^{2}-\frac{3}{4\mu -1}%
(u-v)^{2}].
\end{equation*}%
The second Hamiltonian structure is defined by the Hamiltonian function $%
H^{(1)}=u+v+w$ and the Poisson tensor 
\begin{eqnarray*}
P_{s2}(\mathbf{u}) &=&-\frac{c}{12}\frac{1}{u-v}(\frac{v-w}{w-u}+\frac{w-u}{%
v-w}-2)\left( 
\begin{array}{ccc}
0 & 2 & 1 \\ 
-2 & 0 & -1 \\ 
-1 & 1 & 0%
\end{array}%
\right)  \\
&&+\frac{c}{12}\frac{3\mu }{4\mu -1}\frac{u-v}{(v-w)(w-u)}\left( 
\begin{array}{ccc}
0 & 2 & 1 \\ 
-2 & 0 & -1 \\ 
-1 & 1 & 0%
\end{array}%
\right)  \\
&&+\frac{c}{4}\frac{1-\mu }{4\mu -1}[\frac{1}{w-u}-\frac{1}{v-w}]\left( 
\begin{array}{ccc}
0 & 0 & -1 \\ 
0 & 0 & -1 \\ 
1 & 1 & 0%
\end{array}%
\right) \text{\ .}
\end{eqnarray*}
\end{proposition}

Note that, in the limit $\mu =1/3$ ( or $a=c$) both $\phi _{s}(\mathbf{u})$
(defined as the multiplicative factor in $P_{1}(\mathbf{u})$) and $H_{s}(%
\mathbf{u})$ reduce to some constant multiples of $\phi _{f}(\mathbf{u})$
and $H_{f}(\mathbf{u})$, respectively.

\subsection{Non-interacting two body case}

We consider $a=b\neq 0$ and $c=0$, that is, $\mu =1/4$. This case
corresponds to a situation where two of the three bodies do not interact
with each other but with the third one only. Or, to a situation where the
distance $|u-v|$ is so large that the interaction between bodies at $u$ and $%
v$ can be neglected. Setting the only constant $a=1$, we have

\begin{proposition}
Aristotelian equations of motion%
\begin{equation*}
\acute{u}=\frac{1}{u-w}\text{, \ \ }\acute{v}=\frac{1}{v-w}\text{, \ \ \ }%
\acute{w}=\frac{1}{w-u}+\frac{1}{w-v}
\end{equation*}%
for two bodies at positions $u$ and $v$ interacting with a third one at $w$
are bi-Hamiltonian with%
\begin{eqnarray*}
P_{n1}(\mathbf{u}) &=&\frac{2}{(u-v)^{2}(v-w)(w-u)}\left( 
\begin{array}{ccc}
0 & 1 & -1 \\ 
-1 & 0 & 1 \\ 
1 & -1 & 0%
\end{array}%
\right) \text{, \ \ \ \ \ \ } \\
H_{n}(\mathbf{u}) &=&\frac{1}{4}(u+v-2w)(u-v)^{3} \\
P_{n2}(\mathbf{u}) &=&\text{\ }\frac{1}{6}\frac{u-v}{(v-w)(w-u)}\left( 
\begin{array}{ccc}
0 & 2 & 1 \\ 
-2 & 0 & -1 \\ 
-1 & 1 & 0%
\end{array}%
\right)  \\
&&+\frac{1}{2}[\frac{1}{w-u}-\frac{1}{v-w}]\left( 
\begin{array}{ccc}
0 & 0 & -1 \\ 
0 & 0 & -1 \\ 
1 & 1 & 0%
\end{array}%
\right) , \\
H^{(1)}(\mathbf{u}) &=&u+v+w.
\end{eqnarray*}
\end{proposition}

There remains the case of non-interacting two bodies each of which interact
with the third one with different coupling constants $a,b$, $a\neq b$. We
will give the form of Hamiltonian function of this general non-interacting
two body case in the next section.

\section{General Form of Hamiltonian Function}

We have seen that casting the dynamical system modelling Aristotelian motion
of three bodies into bi-Hamiltonian form requires first the integration of
Eq.(\ref{integ}) to find the time-independent Hamiltonian function and then,
the invariant volume density. In this section, we shall discuss the general
form of characteristics and their domain of validity. Our discussion will,
by no means, be exhaustive.

We first give the form of Hamiltonian function for motions of three bodies
two of which are not interacting with each other. The remaining third body
interacts with them with different coupling constants $a$ and $b$.
Integration in Eq.(\ref{integ}) gives

\begin{proposition}
Let two non-interacting bodies at positions $u$ and $v$ interact with the
third one at $w$. Then, the Aristotelian equations of motion 
\begin{equation*}
\acute{u}=\frac{b}{u-w}\text{, \ \ }\acute{v}=\frac{a}{v-w}\text{, \ \ \ }%
\acute{w}=\frac{b}{w-u}+\frac{a}{w-v}
\end{equation*}%
admit the conserved Hamiltonians $H^{(1)}$ and   
\begin{eqnarray}
H(\mathbf{u}) &=&2\sqrt{4k^{2}-1}\ln \frac{u-v}{\sqrt{2}}  \notag \\
&&+(\sqrt{4k^{2}-1}-k)\ln (\frac{u+v-2w}{\sqrt{3}(u-v)}-\sqrt{4k^{2}-1}+2k)
\\
&&+(\sqrt{4k^{2}-1}+k)\ln (\frac{u+v-2w}{\sqrt{3}(u-v)}+\sqrt{4k^{2}-1}+2k) 
\notag
\end{eqnarray}%
where we introduce the constant $k=(a+b)/(\sqrt{3}(a-b))$.
\end{proposition}

\bigskip Returning to the general case, we first cast Eq.(\ref{integ}) into
the form 
\begin{equation}
h(\mathbf{u})=\ln \eta +\int \frac{(\theta -\theta _{+})(\theta -\theta _{-})%
}{(\theta -\theta _{1})(\theta -\theta _{2})(\theta -\theta _{3})}d\theta 
\label{hamdec}
\end{equation}%
and, after some manipulations, obtain

\begin{proposition}
Let $\theta _{\pm }$ be the roots of quadratic term in numerator and, let $%
\theta _{i}$, $i=1,2,3$ be the roots of cubic polynomial in denominator of
the integral in Eq.(\ref{integ}). Then, the second fundamental
(time-independent) conserved quantity for Aristotelian model of three body
motion is%
\begin{eqnarray}
h(\mathbf{u}) &=&(\theta _{1}-\theta _{2})(\theta _{2}-\theta _{3})(\theta
_{3}-\theta _{1})\ln \frac{u-v}{\sqrt{2}}  \notag \\
&&+(\theta _{1}-\theta _{+})(\theta _{2}-\theta _{3})(\theta _{1}-\theta
_{-})\ln (\frac{u+v-2w}{\sqrt{3}(u-v)}-\theta _{1})  \notag \\
&&+(\theta _{2}-\theta _{+})(\theta _{2}-\theta _{-})(\theta _{3}-\theta
_{1})\ln (\frac{u+v-2w}{\sqrt{3}(u-v)}-\theta _{2})  \notag \\
&&+(\theta _{1}-\theta _{2})(\theta _{3}-\theta _{+})(\theta _{3}-\theta
_{-})\ln (\frac{u+v-2w}{\sqrt{3}(u-v)}-\theta _{3}).  \label{hh3}
\end{eqnarray}
\end{proposition}

To this end, we want to discuss examples of restrictions on the domain of
definition of the fundamental Hamiltonian function $h$. The integral in Eq.(%
\ref{hamdec}) is singular for a real root $\theta _{\func{real}}$ of the
cubic polynomial in denominator. That is, for the case $\theta =\xi /\eta
=\theta _{\func{real}}$. In the real variables $(u,v,w)$ this implies%
\begin{equation*}
u(1-\sqrt{3}\theta _{\func{real}})+v(1+\sqrt{3}\theta _{\func{real}})-2w=0.
\end{equation*}%
The cases described by the conditions $u=v$ (or $\eta =0$), $v=w$ and $w=u$
are particularly included in this equation. Moreover, for non-zero values of 
$\theta _{\func{real}}$, by adding and subtracting $u$ and $v$ to the above
equation we can obtain the singular cases%
\begin{eqnarray*}
v-w &=&0\text{, \ \ \ \ }1-\sqrt{3}\theta _{\func{real}}=0 \\
w-u &=&0\text{, \ \ \ \ }1+\sqrt{3}\theta _{\func{real}}=0
\end{eqnarray*}%
which let the variables $u$ and $v$, respectively, be free but bring
restrictions on the coupling constants $a,b,c$. Finally, the present form of
roots of cubic equation imposes some conditions on the coupling constants
related to the discriminant of cubic polynomial.

\begin{proposition}
For the values of the vector field $\mathbf{u}$ not perpendicular to the
constant vector%
\begin{equation*}
(1-\sqrt{3}\theta _{\func{real}},1+\sqrt{3}\theta _{\func{real}},-2),
\end{equation*}%
the dynamical system in Eq.(\ref{c2}) admits bi-Hamiltonian structure with
Hamiltonian functions in Eqs.(\ref{hh1}) and (\ref{hh3}).
\end{proposition}

In the particular value $\theta _{\func{real}}=0$ of the real root, that may
occur for example in semi- and full-symmetrical configurations, the
condition of above proposition simply prevents the adapted coordinate $\xi $
to become zero. For generic values of $\theta _{\func{real}}$ it just
requires us to be away from the line $\xi =\theta _{\func{real}}\eta $ on
level sets of $H^{(1)}$.

The discussion on domain of validity of bi-Hamiltonian structure will become
more conclusive if we adapt the parameters%
\begin{equation*}
\vartheta =\theta -\frac{p}{3}\text{, \ \ \ \ }p=\frac{a-b}{\sqrt{3}c}\text{%
, \ \ \ \ }q=\frac{a+b}{3c}
\end{equation*}%
for the integration of Eq.(\ref{integ}). In these parameters, the
semi-symmetrical case is characterized by 
\begin{equation*}
p=0\text{, \ \ \ \ \ }q=\frac{2a}{3c}=\frac{1}{3}\frac{1-\mu }{4\mu -1}
\end{equation*}%
and for the full-symmetrical case we have $p=0$, $q=2/3$. Note that, the
non-interactive two body case ($c=0$) is necessarily excluded from the
present discussion. The integral for the fundamental conserved quantity
takes the form%
\begin{equation}
h(\xi ,\eta )=\ln \eta +\int \frac{-(2p^{2}+9q+3)/9-(p/3)\vartheta
+\vartheta ^{2}}{-2p(p^{2}+18q+15)/27-((p^{2}+12q+1)/3)\vartheta +\vartheta
^{3}}d\vartheta   \label{cubic}
\end{equation}%
which can be put into the form of Eq.(\ref{hamdec}) with the roots 
\begin{eqnarray*}
\vartheta _{1} &=&-\frac{1}{3\lambda ^{1/3}}(1+p^{2}+12q+\lambda ^{2/3}) \\
\vartheta _{\pm } &=&\frac{1}{6\lambda ^{1/3}}(\varsigma _{\pm
}(1+p^{2}+12q)+\varsigma _{\mp }\lambda ^{2/3})
\end{eqnarray*}%
of the cubic denominator. Here, $\varsigma _{\pm }=1\pm i\sqrt{3}$ and 
\begin{equation*}
\lambda =-p(p^{2}+18q+15)+\sqrt{27p^{4}+6p^{2}(6(13-3q)q+37)-(1+12q)^{3}}.
\end{equation*}%
In the semi-symmetrical case, we have $\lambda =(-1-12q)^{3/2}$, $q=2a/3c$
and so $\vartheta _{1}=0,$ $\vartheta _{\pm }=\sqrt{1+12q}/\sqrt{3}.$In the
full-symmetrical case, $q=2/3$, $\lambda =27i$ and $\vartheta _{1}=0$, $\ \
\vartheta _{\pm }=\sqrt{3}$. 

As a final example of this analysis, we want to present the connection
between values of the discriminant and the constants appearing in the
integral of Eq.(\ref{integ}). Recall the discriminant 
\begin{equation}
\bigtriangleup =4(-27p^{4}+6p^{2}(-37+6q(-13+3q))+(1+12q)^{3})/27
\end{equation}%
of cubic polynomial in Eq.(\ref{cubic}) which is the negative of the
square-rooted term in $\lambda $. Regarding the right hand side as a
quadratic polynomial in $p^{2}$, we find its discriminant to be%
\begin{equation}
-16777216(1+3q)^{2}(7+3q)^{6}(1+12q)^{3}/177147.  \label{denp}
\end{equation}%
$\bigtriangleup $ will be a perfect square at the values of $%
q=-1/3,-3/7,-1/12$ and these correspond to the following sets of values of
the coupling constants%
\begin{equation*}
a+b+c=0\text{, \ \ \ }7(a+b)+9c=0\text{, \ \ \ \ }4(a+b)+c=0
\end{equation*}%
two of which are seen in Eq.(\ref{integ}). First two restrictions imply, for
the semi-symmetrical motion, $\mu =0$ and $\mu =1/29$, respectively, the
last one, however, is not possible.

We expect the results of the present work to be useful in an exhaustive
analysis of various cases of Aristotelian model of three body motion as well
as in understanding the physics behind this model.

\section{Acknowledgement}

We dedicate this work to the memory of our mentor Professor Yavuz Nutku
(1942-2010). His guidence over the years have always led both of us to
master various geometric aspects of dynamical systems. This work was
initiated by his correspondence, that essentially containing section (\ref%
{nutku}), to one of us (HG). \

\bigskip

\bigskip

\bigskip

\bigskip

\end{document}